\documentclass[aps,pra,superscriptaddress,epsfig,twocolumn]{revtex4-1}

\usepackage{amsmath}
\usepackage{amsfonts}
\usepackage{amssymb}
\usepackage{amstext}
\usepackage{amsthm}
\usepackage{graphicx}
\usepackage{graphics}
\usepackage{times}
\usepackage{bm}
\usepackage{color}
\usepackage{latexsym}
\usepackage{hyperref}
\usepackage{subfigure}
\usepackage{textcomp}
\usepackage{braket}
\usepackage{float}
\usepackage{palatino}
\usepackage{bbm}
\usepackage{tikz}

{\catcode`\|=\active 
 \gdef\Braket#1{\begingroup
\mathcode`\|32768\let\vert\BraVert\left<{#1}\right>\endgroup}}
\def\BraVert{\egroup\,\mid\,\bgroup}

\begin{document}

\title{Kolmogorov-Sinai entropy and dissipation in driven classical Hamiltonian systems}

\author{Matheus Capela}
\email{mthscp27@gmail.com}
\affiliation{Institute of Physics, Federal University of Goi\'{a}s, 74690-900 Goi\^{a}nia, Brazil}

\author{Mikel Sanz}
\email{mikel.sanz@ehu.es}
\affiliation{Department of Physical Chemistry, University of the Basque Country UPV/EHU, Apartado 644, E-48080 Bilbao, Spain}

\author{Enrique Solano}
\email{enr.solano@gmail.com}
\affiliation{Department of Physical Chemistry, University of the Basque Country UPV/EHU, Apartado 644, E-48080 Bilbao, Spain}
\affiliation{IKERBASQUE, Basque Foundation for Science, Maria Diaz de Haro 3, E-48013 Bilbao, Spain}
\affiliation{Department of Physics, Shanghai University, 200444 Shanghai, China}

\author{Lucas C. C\'{e}leri}
\email{lucas@chibebe.org}
\affiliation{Institute of Physics, Federal University of Goi\'{a}s, 74690-900 Goi\^{a}nia, Brazil}

\begin{abstract}
A central concept in the connection between physics and information theory is entropy, which represents the amount of information extracted from the system by the observer performing measurements in an experiment. Indeed, Jaynes' principle of maximum entropy allows to establish the connection between entropy in statistical mechanics and information entropy. In this sense, the dissipated energy in a classical Hamiltonian process, known as the thermodynamic entropy production, is connected to the relative entropy between the forward and backward probability densities. Recently, it was revealed that energetic inefficiency and model inefficiency, defined as the difference in mutual information that the system state shares with the future and past environmental variables, are equivalent concepts in Markovian processes. As a consequence, the question about a possible connection between model unpredictability and energetic inefficiency in the framework of classical physics emerges. Here, we address this question by connecting the concepts of random behavior of a classical Hamiltonian system, the Kolmogorov-Sinai entropy, with its energetic inefficiency, the dissipated work. This approach allows us to provide meaningful interpretations of information concepts in terms of thermodynamic quantities.
\end{abstract}

\maketitle

\section{Introduction}

Dissipation, which occurs when systems are driven out-of-equilibrium, is a fundamental subject in physics, since it is related to reversibility of physical processes. Significant developments has been achieved in the description of such systems, specially concerning their energetics \cite{Jarzynski,Jarzynski1,Evans,Crooks}. Since the development of classical information theory \cite{shannon} and Jaynes' principle of maximum entropy \cite{jaynes} ---the equilibrium probability distribution maximizes information transfer in the measurement process---, several deep links between information theory and thermodynamics have been discovered \cite{john}. 

Based on the observation that logical irreversibility implies thermodynamic irreversibility \cite{landauer}, a relation between the energy cost of computation, which is a physical process, and the algorithmic complexity was derived \cite{zurek}. The algorithmic complexity of a system is defined as the shortest algorithm, measured in bits, whose output is the actual physical configuration of the system. Thus, it is a natural measure of randomness, deeply linked with the information metric, i.e. distance between strings of bits. 

An important relation between dissipation and informational entropy was obtained in Ref. \cite{kawai}. Considering a driven Hamiltonian system, initially in equilibrium, it was proven that the dissipated work due to the driven process is proportional to the Kullback-Liebler divergence (relative entropy) between the forward and backward probability densities ---backward process defined by the time-reverse driven protocol. This result directly leads to the second law of thermodynamics, since relative entropy is a positive semi-definite quantity.
 
Another recent work has revealed that energetic inefficiency, i.e. the dissipated energy, and model inefficiency (non-predictive power) are equivalent concepts in Markovian processes \cite{still}. There, predictive power is the information that the system retains from the past and that is actually necessary to predict the system future behavior \cite{jeffreys,still1,still2}. Therefore, non-predictive information is just the difference between all the information that the systems has and the predictive power.

Using the mathematical formulation provided by Shannon \cite{shannon}, Kolmogorov constructed a theoretical tool, currently known as Kolmogorov-Sinai entropy (KSE), allowing to analyze the random behavior of dynamical systems. It is a parameter of the dynamical system which provides a criterion to define chaos, since positive KSE is a feature of chaotic behavior \cite{frigg}. Despite such construction, the connection between thermodynamic entropy ---or even the informational one--- and KSE for general systems remains unclear \cite{latora,bologna,massimo}. 

This raises the question whether randomness and energetic inefficiency are connected in the framework of classical physics. The purpose of this work is to derive a mathematical relation between the concept of random behavior of a classical Hamiltonian system, measured by the KSE, and its energetic inefficiency, measured by dissipated energy, in accordance with the fluctuation relations. This connection is built by describing a dynamical system in terms of communication theory.  is unambiguous, it paves the way for a novel route to investigate quantities like work and heat in the framework of quantum thermodynamics.

The manuscript is organized as follows: we start by defining the class of systems under consideration, followed by a description of an informational-theoretic approach for dynamical systems, in which the concept KSE of the dynamics is introduced. Then, in the main part of the article, we present our result, which establishes a lower bound on the dissipation in terms of the randomness generated by the dynamics. We close the paper with a proper discussion regarding the application of the result in classical systems. 

\section{Physical setup}

The system is described by a Hamiltonian $H\left(s_t;\lambda\right)$, with $s_t =\left(q(t),\theta(t)\right)$ the set of generalized coordinates and canonical conjugate momenta in phase space $\Gamma$. The control parameter $\lambda$ is varied in time following an externally controlled protocol $\lambda(t)$, denoted the {\it work protocol}. We assume that the system is initially in thermal equilibrium with a reservoir at inverse temperature $\beta$, thus the distribution function is given by
\begin{equation}
\rho_0(s_{0},\lambda_{0}) = \frac{e^{-\beta H(s_{0};\lambda_{0}) } }{Z(\lambda_0)},
\end{equation}
where $Z(\lambda)=\int_{\Gamma} ds\exp\{ -\beta H(s;\lambda) \}$ is the partition function and $\lambda_0 \equiv \lambda(0)$. During the action of the work protocol, the system is isolated from the reservoir. This means that the energy exchange between the system and the external world is determined by $\lambda$, and we denote this energy as the work $W$.
 
The dynamics in phase space is deterministic and governed by the Hamilton equations
\begin{eqnarray}
\dot{q_i} & = & \frac{ \partial H(s;\lambda)}{\partial \theta_i}, \nonumber \\
\dot{\theta_i} & = & -\frac{ \partial H(s;\lambda)}{\partial q_i}. \nonumber
\end{eqnarray}
We denote the Hamiltonian flow, i.e., the time evolution map that determines the trajectory $s_t$ associated with each initial condition $s_0$, by $s_t=\phi^t(s_0)$. Therefore, our dynamical system is the triplet $(\Gamma,p,\phi^t)$, a probability space, $(\Gamma,p)$, equipped with a one-parameter group of automorphisms of the probability measure space, $\phi^t$, with each time evolution function $\phi^t$ depending on time~$t$ \cite{arnold,sinai,frigg}. Additionally, $p : \Sigma \rightarrow [0,1]$ is the initial probability measure over the sigma-algebra $\Sigma$. For Hamiltonian systems, $\Sigma$ is simply the set of all the subsets of the phase space $\Gamma$. Finally, we would like to highlight that, due to the Hamiltonian dynamics, Liouville's theorem applies.

Let us now define, for every time $t$, the functional 
\begin{equation}
S[\rho_t] = -\int \limits_{\Gamma} ds \rho_t(s)\ln \rho_t(s),
\end{equation}
which is the Shannon differential entropy associated to the density $\rho_t(s)$ at time $t$ \cite{cover}. We observe that $S$ is defined considering the support of the probability distribution. Initially, as canonical equilibrium is assumed, this is equal to the thermodynamic entropy and thus, contains all the thermodynamic information about the system. However, in general, this is no longer true for $t>0$, since the work protocol acting on the system may lead it out of equilibrium.

\section{Randomness}

The main goal of the present study is to establish a connection between the dissipated energy during the work protocol and the Kolmogorov-Sinai entropy (KSE). Therefore, as we must define this entropy for the dynamical system, we should seek an information-based description of the dynamical system. Usually, in a communication process, the source emits to the receiver discrete symbols (the messages) according to certain probability distribution. The KSE quantifies the randomness of this process. The goal is to compute it for a dynamical system and associate randomness to chaos. Consequently, in order to relate KSE to a dynamical system, we need to define a discrete alphabet, which is accomplished by partitioning the phase space $\Gamma$ \cite{frigg}.

We say that a collection $A$ of subsets is a partition of the phase space $\Gamma$, if its elements $\alpha \in A$ are disjoint, i.e. $\forall\,\,\, \alpha,\alpha' \in A$, $\alpha \cap \alpha' = \emptyset$ if $\alpha \neq \alpha'$, and cover the $\Gamma$, $\bigcup_{\alpha \in A}\alpha =\Gamma$. Let us remark that, from two given partitions, $A$ and $B$, it is possible to define a new partition, $A \lor B$, by means of the {\it refinement} $A \lor B  = \left\lbrace\alpha \cap \beta \,|\, \alpha \in A \, , \, \beta \in B\right\rbrace$.

Let us now focus on the Hamiltonian flow. In order to simplify the analysis, we will consider a discrete version of the dynamical system. This means that the time $t$ is discrete, $t \in \mathbb{Z}$, and the time evolution is generated by the iteration of the automorphism $\phi \equiv \phi^1$. Let us remark that such assumption is physically sensible and reasonable, since it is in accordance with the statement that any measurement is discrete in time from an experimental point of view. In this context, time evolution is introduced by means of the refinement of partitions as follows. Let us consider some initial partition $A$, whose elements are denoted by $\alpha$. Then, an {\it evolved} partition $\phi(A)$ is obtained by the application of the evolution map $\phi$ on $A$, i.e. $\phi(A) = \left\lbrace\phi(\alpha) \,|\, \alpha \in A\right\rbrace$. Therefore, trajectories in phase space provide us with a discrete sequence of partitions $A, \phi(A), \phi^{2}(A),\ldots$, based on an initial partition $A$. This is the required discrete alphabet to extend the definition of KSE for  dynamical systems \cite{frigg}.

From these definitions, the entropy of a finite partition $A$ with respect to the probability measure $p$ can be defined as the Shannon entropy of the probability vector $[p(\alpha)]_{\alpha \in A}$
\begin{equation}
S(A) = \sum_{\alpha \in A} z [p(\alpha)],
\end{equation}
where $z(x)=-x\ln x$ if $x>0$, and $z(x)=0$ if $x=0$. We can interpret this entropy as the amount of uncertainty concerning the element of the partition in which the state is. 

We are now ready to define the randomness of the dynamical system. Let us define the KSE associated with the dynamics as~\cite{arnold}
\begin{equation}
 h(\phi) := \sup_{A \in P} \lim_{t \rightarrow \infty} \frac{S \left[\bigvee_{n=0}^{t-1}\phi^{-n}(A)\right]}{t},
 \label{eq:kse}
\end{equation}
where $\bigvee_{n=0}^{t-1}\phi^{-n}(A)=A \lor \phi^{-1}(A) \lor \dots \lor \phi^{-t+1}(A)$ and $P$ is the set of all finite partitions of the phase space. For the sake of clarity, we have denoted as $\phi^{-n}$ the application of the inverse map $(\phi^{-1})^n$. Indeed, note that the Hamiltonian flux $\phi$ is a bijection and hence, there exists its inverse. Consequently, we can compute all the transition probabilities only relying on the initial probability $p_0$. If $p_t$ represents the probability measure at time $t$, then $p_t[A \lor \dots \lor \phi^{t-1}(A)] = p_0[\phi^{-t+1}(A)\lor\dots\lor A] = p[A \lor \dots \lor \phi^{-t+1}(A)]$. This implies that $S_{t} \left[A \lor \dots \lor \phi^{t-1}(A)\right] = S \left[\bigvee_{n=0}^{t-1}\phi^{-n}(A)\right]$ and Eq. (\ref{eq:kse}) follows, where $S_{t}$ is the Shannon entropy computed at time $t$.

Starting from some initial condition $s_0$, we can follow the associated orbit and, at every instant of time $t>0$, assigning an element of the partition, $\alpha_{t} \in A$, which contains the state $s_t$. In other words, for every initial condition and every partition, we can construct the path $(\alpha_{0},\alpha_{1},\dots,\alpha_{t})$. Note that $\alpha_t$ is an element of the original partition $A$ in which the system is at time $t$, it does not necessarily mean that $\alpha_t$ and $\alpha_{t+1}$ are contiguous. This is done in order to avoid the introduction of an extra index to label the elements of the partition. 

Given the above definitions, we are interested in the probability of observing a specific path $(\alpha_{0},\alpha_{1},\dots,\alpha_{t})$. According to the theory of dynamical systems, such probability is defined as \cite{frigg,sinai}
\begin{equation}
 p(\alpha_{0},\alpha_{1},\dots,\alpha_{t}) = p[\alpha_{0} \cap \phi^{-1}(\alpha_1) \cap \dots \cap \phi^{-t}(\alpha_t)],
\end{equation}
from which the conditional probabilities can be calculated as
\begin{equation}
 p(\alpha_{t}|\alpha_{0},\alpha_{1},\dots,\alpha_{t-1})=\frac{p(\alpha_{0},\alpha_{1},\dots,\alpha_{t})}{p(\alpha_{0},\alpha_{1},\dots,\alpha_{t-1})}.
\end{equation}

The KSE is a standard tool in the analysis of chaotic behavior, in a sense that it quantifies the notion of randomness in the coarse-grained phase-space dynamics. We can understand this statement by looking at the equation
\begin{align}
h(\phi) = -\sup_{A \in P} & \lim_{n \rightarrow \infty} \frac{1}{n} \sum_{t=1}^n \sum_{\alpha_0, \dots, \alpha_{t-1}} p(\alpha_{0},\dots, \alpha_{t-1}) \nonumber \\
&\times \sum_{\alpha_t} z[p(\alpha_{t}|\alpha_{0},\dots, \alpha_{t-1})].
\label{eq:KSE}
\end{align}
The RHS of the above equation is known as the generalized Shannon entropy of the dynamical system. It measures our knowledge about the dynamics of the system \cite{frigg}. Indeed, the higher the uncertainty about the dynamics, the larger the randomness generated by it. 

\section{A lower bound on dissipation}

It is known that coarse-graining increases informational entropy, since we are effectively discarding information \cite{benatti,ana}. Here, we are interested in studying the thermodynamic entropy production rate during the dynamical evolution of the system. We start by defining the conditional coarse-grained density of states in the phase space as
\begin{equation} \label{density}
\rho^{cg} (s|\alpha_{0},\dots, \alpha_{t-1}) = \sum_{\alpha_t \in A} \frac{p(\alpha_t|\alpha_{0},\dots, \alpha_{t-1})}{v(\alpha_t)} \mathbbm{1}_{\alpha_t}(s),
\end{equation}
where we have defined the indicator function $\mathbbm{1}_{\alpha}(s) = 1$ if $s\in\alpha$ and $\mathbbm{1}_{\alpha}(s) = 0$ otherwise, and the phase space volume $v(\alpha)=\int_{\alpha}ds$.

We focus now on our main objective, which is to connect the randomness generated by the dynamics with the dissipated work (thermodynamic entropy production). We know that the entropy production is related to an entropic quantity by means of the fluctuation relation \cite{kawai}. Therefore, we need to establish the connection between this entropic quantity and the KSE given by Eq. (\ref{eq:KSE}). In order to achieve this goal, we will compute the averaged Shannon entropy of the density of states in phase space, given by Eq. (\ref{density}), $\mathbb{E}_{p(\alpha_{0},\dots, \alpha_{t-1})}S[\rho^{cg}(\alpha_{0},\dots, \alpha_{t-1})]:=\mathbb{E}S[\rho_t^{cg}]$, which is lower bonded by (see Appendix \ref{Appendix} for details)
\begin{equation}
\mathbb{E}S[\rho_t^{cg}] \geq S[\rho_t] + c_t(A) + d_t(A),
\label{eq:lower}
\end{equation}
where 
\begin{align}\label{c}
 c_t(A) = & 1-\sum_{\alpha_0, \dots,  \alpha_{t} \in A} p(\alpha_{t}|\alpha_{0},\dots, \alpha_{t-1})\times \nonumber \\
 & \times \tilde{v}(\alpha_{t-1}, \alpha_{t-2}, \dots, \alpha_{0} | \alpha_{t}), 
\end{align}
and
\begin{equation}
 d_t(A) := -\sum_{\alpha_t\in A} p[\phi^{-t}(\alpha_t)] \ln v[\alpha_t].
\end{equation}
We note that we defined $p(\alpha_0|\alpha_0) := p(\alpha_0)$ and $v(\alpha_0|\alpha_0) := 1$. The renaming terms ($t>1$) of $p(\alpha_{t}|\alpha_{0},\dots, \alpha_{t-1})$ and $v(\alpha_{t}|\alpha_{0},\dots, \alpha_{t-1})$ are defined as usual.

We note that the tilde over any quantity means that we are looking at the time-reversed trajectory. 

In the present setup, the dissipated work is given by $\langle W_d\rangle = \beta^{-1}S(\rho||\tilde{\rho})$, with $S(\cdot||\cdot)$ denoting the relative entropy between its arguments, while $\langle\cdot\rangle$ stands for phase space average. Now, by taking the time average of this equation and also of Eq. (\ref{eq:lower}), we can relate the obtained results with the KSE through Eq. (\ref{eq:KSE}) (see Appendix \ref{Appendix} for details). The result is, hence,
\begin{equation}\label{result}
 \beta \overline{ \langle W_{d} \rangle }\geq \beta ( \overline{ \langle H \rangle }-  \overline{F(\lambda_t)}) - \overline{I}_{t}(A).
\end{equation}
In this equation, we define $F(\lambda_t):=\beta^{-1}\ln Z(\lambda_t)$ as the reference free energy at time $t$ and $ \overline{I_{t}}(A) = h(\phi) - \overline{ c_t(A) }- \overline{ d_t(A)}$. Equation (\ref{result}) is the main result of this work.

\subsection{Examples}

Let us consider as a first example of dynamical system the rotation of the unit disk in the two dimensional phase space. The Hamiltonian function that generates such dynamics is the quadratic form $H(q,\theta)=q^2+\theta^2$, where the dimensionless energy is constrained to be smaller than one. This is just a harmonic oscillator. Setting the unit of time as the time spent in a rotation of $\pi$ radians and considering the trivial partition $A=\{\Gamma\}$, the quantity $\overline{I}_{t}(A)$ is equal to zero, since $h(\phi)=0$ for integrable dynamics. The bound is then saturated in the infinite temperature limit. If we consider a different partition $A=\{[q>0], [q\leq 0]\}$, where $[q>0]$ ($[q\leq 0]$) represents the states in the unit disk with positive coordinate $q$ (non-positive coordinate $q$), we have $\overline{I}_{t}(A) = 1 - ln 2$, thus implying in a negative, trivial bound. This simple example highlight the fact that the result is dependent on the chosen partition, which is expected since KSE is partition dependent, being maximal for the generating one. This is the case in which our result is not informative, since we do not have external driven and the dissipated work is zero.

Let us now move to a more interesting example by considering the case of a chaotic system, the paradigmatic kicked top. The system Hamiltonian can be written in the form $H = \alpha J_{x} + \kappa J_{z}^{2}\sum_n \delta(t-n)$, where $J_{i}$ is $i$-th component of the angular momentum and we choose a unit period of the kicks. It is easy to see that the total angular momentum is conserved, thus restricting the phase space to a sphere. Choosing the energy scale appropriately we have that the system is chaotic for $\kappa \gtrsim 2$, in the sense that the greatest Lyapunov exponent becomes positive. We are interested here in the complete chaotic regime, in which the entire phase space is chaotic. This happens for $\kappa \gtrsim 5$. 

In this regime we can compute the time averages that appears in Eq. (\ref{result}) by considering the fact that the system quickly thermalize, i.e. all points in phase space have the same probability. Let us start with $\overline{ c_t(A) }$, which is the most difficult quantity to numerically compute. In our case the trajectories of the system would be so sparse that, for any sequence $\alpha_0, \dots, \alpha_{t-1}$, the system will assume equal probabilities for the event $\alpha_t$. Mathematically   $p(\alpha_{t}|\alpha_{0},\dots, \alpha_{t-1})=1/n$, where $n$ is the number of elements in the considered partition. Due to the same reason, the conditional volume reduces simply to $ \tilde{v}(\alpha_{t-1}, \alpha_{t-2}, \dots, \alpha_{0} | \alpha_{t})= (1/n)^t$. Plugging this into Eq. (\ref{c}) results in $\overline{ c_t(A) }=0$ for the complete chaotic case.

Regarding the calculation of the quantity $\overline{d_t(A)}$, consider the following reasoning, which is similar to the one used above for $\overline{ c_t(A) }$. First, let us point out that $p[\phi^{-t}(\alpha_t)]$ is just the probability of finding the system state in $\alpha_t$ at time $t$, as implied by Liouville's theorem. Since the dynamics is chaotic, there is an instant of time $t$ that the probability $p[\phi^{-t}(\alpha_t)]$ is approximately equal to the volume $v(\alpha_t)/v(\Gamma)$. Then, the quantity $d_t(A)$ is approximately equal to $-\sum_\alpha [v(\alpha)/v(\Gamma)]\ln v(\alpha)$. If $A$ is a partition such that its cells have equal volume the quantity  $\overline{ d_t(A) }$ reduces to $-n [v(\alpha)/v(\Gamma)]\ln v(\alpha)$. Note that if $n$ tends to infinity, then $v(\alpha)/v(\Gamma)$ tends to zero. Therefore, Eq. (\ref{result}) reduces to
\begin{equation}
\beta \overline{ \langle W_{d} \rangle }\geq \beta ( \overline{ \langle H \rangle }-  \overline{F(\lambda_t)}) - h(\phi) + n [v(\alpha)/v(\Gamma)]\ln v(\alpha).
\end{equation}
We can compute the value of KSE by means of Pesin identity. The for the case $\kappa = 5$ (and $\alpha = \pi/2$) is $h(\phi) = 0.8768$.

Regarding the saturation of our result, we note that the inequality appearing in Eq. (\ref{result}) of the main text is a consequence of the approximation considered in Eq. (\ref{eqapx}) of Appendix A. Therefore, inequality (\ref{result}) saturates when the time average of coarse-grained entropy (averaged over all trajectories) is equal to the entropy itself. This implies that the KS entropy must be equal to the initial thermodynamic entropy (Boltzmann constant is equal to 1) per unit time.

\section{Discussion} 

We can understand Eq. (\ref{result}) in the following way. The difference between the Shannon entropies before and after imposing the coarse-graining is called hidden information \cite{ana}. This is the information ignored by coarse-graining the system. Hence, according to this definition, the quantity $d_0(A)$ can be interpreted as the minimum amount of hidden information, since $S[p(\alpha)]-S[\rho_0] \geq d_0(A)$ ($c_0(A)=0$). The equality holds when $\rho_0$ is the uniform distribution. In this way, the quantity $S[p(\alpha)]-d_0(A)$ represents the maximum information that is not hidden, i.e. the information which is not related to the coarse-graining procedure. This is the maximum value which the differential entropy can assume.

Following this line of reasoning, the quantity $\overline{\mathbb{E}_{p(\alpha_{0},\dots, \alpha_{t-1})}S[p(\alpha_t|\alpha_{0},\dots, \alpha_{t-1})]}-\overline{S[\rho_t]} \equiv \overline{I_t^h}$ is naturally interpreted as the average hidden information. Thus, the quantity $\overline{c_t(A)}+\overline{d_t(A)}$ represents the minimum average hidden information, since $\overline{I_t^h}\geq \overline{c_t(A)}+\overline{d_t(A)}$. Therefore, the quantity $\overline{I_{t}}(A)$ represents the maximum average of the information that is not hidden information ---$A$ is the partition that maximizes $\overline{I_t^h}$. This is the information generated by the system dynamics.

The study of out-of-equilibrium systems is in general a very hard task. For instance, the result in Ref.~\cite{kawai} established a powerful connections between information theory and thermodynamics, relating the dissipated work in a Hamiltonian process to the asymmetry between the time evolution of the system and its time-reverse counterpart. The purpose of the present study is to provide new tools for investigating the thermodynamics of out-of-equilibrium Hamiltonian systems. Indeed, based on the understanding of dynamical systems in terms of communication theory, we were able to derive a lower bound on the dissipated work (entropy production) in terms of the complexity of the dynamics, measured by the KSE. In other words, we build a connection between a dynamical quantity, KSE, and a macroscopic physical one, the dissipated work.

In the development of the present work we have assumed that Liouville's theorem applies, which is the case for closed Hamiltonian systems. Based on some results in literature dealing with fluctuation relations for open classical systems \cite{Campisi}, we do believe that our results can be extended to such scenario, including non-Markovian dynamics. However, the connection between communication theory and dynamical maps in this case must be defined very carefully.

Regarding the extension of our work to quantum systems, the author of Ref. \cite{lindblad} defined a dynamical entropy by proposing an adequate definition of quantum stochastic processes. However, we have currently a much deeper understanding of such processes \cite{kavan}, and establishing a connection between them and quantum communication theory should be possible. This path might lead us to an unambiguous definition of KS entropy in the quantum setting.

\vspace{0.2cm}
\noindent
\textbf{Acknowledgments} --- Discussions with I\~{n}igo L. Egusquiza and Kavan Modi are greatly acknowledged. MC and LCC acknowledge financial support from the Brazilian funding agencies CNPq (Grants No. 401230/2014-7, 305086/2013-8 and 445516/2014-), CAPES (Grant No. 6531/2014-08), and the Brazilian National Institute of Science and Technology of Quantum Information (INCT/IQ). M.S. and E.S. are grateful for the funding of Spanish MINECO/FEDER FIS2015-69983-P and Basque Government IT986-16. This material is also based upon work supported by the U.S. Department of Energy, Office of Science, Office of Advance Scientific Computing Research (ASCR), under field work proposal number ERKJ335. LC acknowledges the warm hospitality of QUTIS group at the University of the Basque Country.

\vspace{0.2cm}

\appendix

\section{Derivation of the Lower Bound for the Dissipated Work} \label{Appendix}

The Shannon entropy of $\rho^{cg}(s| \alpha_{0}\dots \alpha_{t-1})$ averaged over all possible paths $(\alpha_{0},\dots,\alpha_{t-1})$ can be compared to the Shannon entropy of $\rho_t$
\begin{widetext}
\begin{eqnarray} \label{eqapx}
\mathbb{E}S[\rho_t^{cg}] - S\left[ \rho_t \right] & = &  \int ds \rho_t(s)\ln \rho_t(s) -  \sum_{\alpha_0, \dots,\ \alpha_{t}} p(\alpha_{0},\dots, \alpha_{t}) \ln \frac{p(\alpha_{t}|\alpha_{0},\dots,\alpha_{t-1})}{v(\alpha_{t})} \nonumber \\
	    & = &  \sum_{\alpha_0, \dots\,  \alpha_{t}} \int_{\alpha_{t} \cap \phi (\alpha_{t-1})\cap \dots \cap \phi^{t}(\alpha_{0}) } ds \rho_t(s)  \left[ \ln \rho_t(s) - \ln \frac{p(\alpha_{t}|\alpha_{0}\dots\alpha_{t-1})}{v(\alpha_{t})} \right] \nonumber \\
	    &\geq& \sum_{\alpha_0, \dots\,  \alpha_{t}} \int_{\alpha_{t} \cap \phi (\alpha_{t-1})\cap \dots \cap \phi^{t}(\alpha_{0})  } ds \bigg [ \rho_t(s) - \frac{p(\alpha_{t}|\alpha_{0}\dots\alpha_{t-1})}{v(\alpha_{t})} \bigg ] \nonumber \\
	    & = & 1-\sum_{\alpha_0, \dots,  \alpha_{t}} p(\alpha_{t}|\alpha_{0}\dots\alpha_{t-1}) \frac{v[\alpha_{t} \cap \phi (\alpha_{t-1})\cap \dots \cap \phi^{t}(\alpha_{0}) ]}{v[\alpha_{t}]},
\end{eqnarray}
\end{widetext}
where the inequality follows from the relation $x(\ln x - \ln y) \geq x-y$, $\forall \, x,y \in \mathbb{R}_+^*$. It is possible to rewrite the quantity $c_t$ considering the volume measure as a time-reversed measure. Since $\psi:=\phi^{-1}$ preserves volume measure due to Liouville's theorem, it follows that

\begin{align}
v[\alpha_{t} \cap \dots \cap \phi^{t}(\alpha_{0})] &= v[\psi^t(\alpha_{t}) \cap \dots \cap \alpha_{0}] \nonumber \\
&= \tilde{v}(\alpha_{t}, \dots, \alpha_{0} ),
\end{align}
which will below lead to Eq. (\ref{c}) for $c_t$. As the time rate of $\mathbb{E} S[\rho_{t}^{cg}] + \sum_{\alpha_t\in A} p[\phi^{-t}(\alpha_t)] \ln v[\alpha_t] $ is equivalent to KSE, when calculated for the finite partition which maximizes it, it follows the lower bound on KSE
\begin{equation}\label{lower}
 h(\phi) \geq \overline{S[\rho_t]} + \overline{c_t(A)} + \overline{d_t(A)},
\end{equation}
where
\begin{align}
c_t(A) :=  1-&\sum_{\alpha_0, \dots,\alpha_{t}} p(\alpha_{t}|\alpha_{0}\dots\alpha_{t-1}) \nonumber \\
	    &\times \frac{v[\alpha_{t} \cap \phi (\alpha_{t-1})\cap \dots \cap \phi^{t}(\alpha_{0}) ]}{v[\alpha_{t}]},
\end{align}
and
\begin{equation}
 d_t(A) := -\sum_{\alpha_t\in A} p[\phi^{-t}(\alpha_t)] \ln v[\alpha_t].
\end{equation}

The free energy is defined as the Legendre transformation $F(\lambda_0):=\langle H \rangle_{\rho_0} -\beta^{-1}S[\rho_0]$. Note that the initial distribution is a equilibrium one and thus, by using Liouville's theorem, it follows the relation $S[\rho_t] = \beta (\langle H \rangle_{\rho_0} - F[x_0]), \, \forall t \in \mathbb{R}$. Additionally, the dissipated work up to time $t$, $\langle W_{d} \rangle_{t} = [\langle H \rangle_{\rho_t} - \langle H \rangle_{\rho_0}] - [F(\lambda_t) -F(\lambda_0)]$ \cite{kawai}, which can be calculated in processes like the ones considered here as
\begin{equation}
 \beta \langle W_{d} \rangle_{t} = -S[\rho_t] + \beta [\langle H \rangle_{\rho_t} - F(\lambda_t)].
\end{equation}
Finally, by solving for $\overline{S[\rho_t]}$ and using inequality (\ref{lower}), we achieve the desired lower bound given by Eq. (\ref{result}).


\begin{thebibliography}{99}

\bibitem{Jarzynski} C. Jarzynski, \emph{Equalities and inequalities: Irreversibility and the second law of thermodynamics at the nanoscale}, Ann. Rev. Condens. Matter \textbf{2}, 329 (2017).

\bibitem{Jarzynski1} C. Jarzynski, \emph{Nonequilibrium equality for free energy differences}, Phys. Rev. Lett. \textbf{78}, 2690 (1997).

\bibitem{Evans} D. J. Evans, E. G. D. Cohen and G. P. Morriss, \emph{Probability of second law violations in shearing steady states}, Phys. Rev. Lett. \textbf{71}, 2401 (1993).

\bibitem{Crooks} G. E. Crooks, \emph{Entropy production fluctuation theorem and the nonequilibrium work relation for free energy differences}, Phys. Rev. E \textbf{60}, 2721 (1999).

\bibitem{shannon} C. E. Shannon, \emph{A mathematical theory of communication}, Bell System Technical Journal \textbf{27}, 379 (1948).

\bibitem{john} J. Goold, M. Huber, A. Riera, L. del Rio and P. Skrzypczyk, \emph{The role of quantum information in thermodynamics ---a topical review}, J. Phys. A: Math. Theor. \textbf{49}, 143001 (2016).

\bibitem{cover} T. M. Cover and J. A. Thomas, \emph{Elements of Information Theory} (Wiley \& Sons, New Jersey, 2006).

\bibitem{jaynes} E. T. Jaynes, \emph{Information theory and statistical mechanics}, Phys. Rev. \textbf{106}, 620 (1957).

\bibitem{still} S. Still, D. A. Sivak, A. J. Bell and G. E. Crooks, \emph{Thermodynamics of prediction}, Phys. Rev. Lett. \textbf{109}, 120604 (2012).

\bibitem{landauer} R. Landauer, \emph{Irreversibility and heat generation in the computing process}, IBM J. Res. Dev. \textbf{3}, 113 (1961).

\bibitem{zurek} W. H. Zurek, \emph{Thermodynamic cost of computation, algorithmic complexity and the information metric}, Nature \textbf{341}, 119 (1989).

\bibitem{kawai} R. Kawai, J. M. R Parrondo and C. Van den Broeck, \emph{Dissipation: The phase-space perspective}, Phys. Rev. Lett. {\bf 98}, 080602 (2007).

\bibitem{arnold} V. I. Arnold and A. Avez, \emph{Ergodic Problems of Classical Mechanics} (WA Benjamin, 1968).

\bibitem{frigg} R. Frigg, \emph{In what sense is the Kolmogorov-Sinai entropy a measure for chaotic behaviour? Bridging the gap between dynamical systems theory and communication theory}, Brit. J. Phil. Sci. \textbf{55}, 411 (2004).

\bibitem{tolman} R. C. Tolman, \emph{The Principles of Statistical Mechanics} (Oxford University Press, 1938).

\bibitem{sinai} I. P. Cornfeld, S. V. Fomin and Ya. G. Sinai, \emph{Ergodic Theory} (Springer-Verlag, 1982).

\bibitem{ruelle} J.-P. Eckmann and D. Ruelle, \emph{Ergodic theory of chaos and strange attractors}, Rev. Mod. Phys.  \textbf{57}, 617 (1985).

\bibitem{benatti} F. Benatti, \emph{Deterministic Chaos in Infinite Quantum Systems} (Springer, 1960).

\bibitem{ana} A. Alonso-Serrano and M. Visser, \emph{Coarse graining Shannon and von Neumann entropies}, Entropy \textbf{19}, 207 (2017).

\bibitem{jeffreys} H. Jeffreys, \emph{Theory of Probability} (Oxford University, New York, 1998).

\bibitem{still1} S. Still, \emph{Information-theoretic approach to interactive learning}, Europhys. Lett. \textbf{85}, 28005 (2009).

\bibitem{still2}  S. Still, J. P. Crutchfield and C. Ellison, \emph{Optimal causal inference: Estimating stored information and approximating causal architecture}, Chaos \textbf{20}, 037111 (2010).

\bibitem{latora} V. Latora and M. Baranger, \emph{Kolmogorov-Sinai entropy rate versus physical entropy}, Phys. Rev. Lett. \textbf{82}, 520 (1999).

\bibitem{bologna} M. Bologna, P. Grigolini, M. Karagiorgis and A. Rosa, \emph{Trajectory versus probability density entropy}, Phys. Rev. E \textbf{64}, 016223 (2001).

\bibitem{massimo} M. Falcioni, L. Palatella and A. Vulpiani, \emph{Production rate of the coarse-grained Gibbs entropy and the Kolmogorov-Sinai entropy: A real connection?}, Phys. Rev. E \textbf{71}, 016118 (2005).

\bibitem{JarzynskiPRX} C. Jarzynski, H. T. Quan and S. Rahav, \emph{Quantum-classical correspondence principle for work distributions}, Phys. Rev. X \textbf{5}, 031038 (2015). 

\bibitem{Campisi} M. Campisi, P. H\"{a}nggi, and P. Talkner. Colloquium: Quantum fluctuation relations: Foundations and applications. Rev. Mod. Phys. \textbf{83}, 771 (2011).

\bibitem{lindblad} G. Lindblad, \emph{Non-Markovian quantum stochastic processes and their entropy}, Comm. Math. Phys. \textbf{65}, 281 (1979).

\bibitem{kavan} F. A. Pollock, C. Rodr\'{i}guez-Rosario, T. Frauenheim, M. Paternostro, and K. Modi. Non-Markovian quantum processes: Complete framework and efficient characterization. Phys. Rev. A \textbf{97}, 012127 (2018).

\end{thebibliography}
\end{document}